\documentclass{appolb}
\usepackage{epsfig}
\usepackage{color}

\begin{document}
\title{Can the massive neutron star PSR J0348+0432 be a hyperon star?}
\author{XIAN-FENG ZHAO$^{1,2}$
\\$^{1}$ School of Sciences, Southwest Petroleum University, \\
{Chengdu, 610500 China}
\\$^{2}$ School of Electronic and Electrical Engineering, \\
{Chuzhou University, Chuzhou 239000, China}
}
\maketitle
\begin{abstract}
Whether the massive neutron star PSR J0348+0432 can change into a hyperon star is studied in the framework of the relative mean field theory by choosing the suitable hyperon coupling constants. We find that whether the mesons $\sigma^{*}$ and $\phi$ being considered or not, the neutron star PSR J0348+0432 all can change into a hyperon star and the hyperon star transition density are the same for the two cases. We also find that the canonical mass neutron star also can change into a hyperon star in a minor hyperon star transition density as the mesons $\sigma^{*}$ and $\phi$ are not considered. Our results confirms some of recent conclusions.

\end{abstract}


\setcounter{page}{171}

\section{Introduction}
When the neutron chemical potential exceeds a certain mass of hyperon, the neutron will be converted to the hyperon. These situations can occur frequently within the neutron stars. In the neutron star, the chemical potentials of the neutrons will increase with the increase of the baryon number density. Once they exceed the mass of a hyperon, the hyperon would produce. In the inner region of the neutron star, the baryon number density is so high that the chemical potentials of the neutrons would be greater than the masses of multiple hyperons, such as $\Lambda$, $\Sigma$ and $\Xi$ et al. As soon as the hyperons produce, we can think the neutron star has changed into a hyperon star~\cite{Glendenning85}.

The critical baryon number density, at which the hyperons produce, is known as the transition density of a hyperon star.

In 2010, a massive neutron star PSR J1614-2230 with the mass 1.97$^{+0.04}_{-0.04}$ M$_{\odot}$ was observed by Demorest et al~\cite{Demorest10}. Recently, its mass was determined to be $1.928^{+0.017}_{-0.017}$ (68.3\% credibility)~\cite{Fonseca16}.

Up to now, the massive neutron stars have been widely studied in various models, such as the massive hadronic neutron stars~\cite{{Zhaoprc12},{Bednarek12},{Bednarek13},{Bejger13},{Jiang12},{Miyatsu13},{Weissenborn12}}, the quark-hadron hybrid stars~\cite{Klahn13,Lastowiecki11,Masuda12}, the hadronic stars within other approximations such as Brueckner-Hartree-Fock~\cite{Massot12,Whittenbury12,Katayama12}.

The recently observed massive neutron star PSR J0348+0432 with the mass of $2.01\pm0.04$ M$_{\odot}$~\cite{Antoniadis13} has aroused people's great interest. For us, the most interesting is the conversion from neutrons to hyperons inside the neutron star. Further more, can the massive neutron star PSR J0348+0432 change into a hyperon star? To solve this problem can tell us the structural information inside the massive neutron star.

In the calculations of the neutron star with the relativistic mean field (RMF) approximation~\cite{{Zhou03},{Zhou10}}, there are two kinds of parameters should be determined.

The first one is the nucleon coupling constant. The results show that the GL85 set can better describe the properties of the neutron stars~\cite{Glendenning85,Glendenning91,Zhaoprc12}.

The another one is the hyperon coupling constant. For this kind of parameters, the hyperon coupling constants coupling to the vector mesons $\omega$ can be chosen by the quark constituent SU(6) symmetry and those coupling to the scalar mesons $\sigma$ by fitting to the $\Lambda, \Sigma$ and $\Xi$ well depth in saturation nuclear matter.

The above model only consider the interactions between the nucleon-nucleon (NN) and the hyperon-nucleon (YN), being described by the mesons $\sigma$, $\omega$ and $\rho$. We name it as $\sigma \omega \rho$ model.

To finely describe the interaction between the hyperon-hyperon (YY), the mesons $f_{0}(975)$ (denoted as $\sigma^{*}$)and $\phi(1020)$ (denoted as $\phi$), which only interact between hyperons, should be considered~\cite{Schaffner94}. This is named as $\sigma \omega \sigma^{*} \phi \rho$ model.

In this paper, we study whether the massive neutron star PSR J0348+0432 can change into a hyperon star in the framework of the RMF theory with the above two models.

\section{The RMF theory and the mass of the massive neutron star PSR J0348+0432}
We start from the Lagrangian density of hadron matter containing mesons $\sigma^{*}$ and $\phi$~\cite{Schaffner94,Glendenning97}
\begin{eqnarray}
\mathcal{L}&=&
\sum_{B}\overline{\Psi}_{B}(i\gamma_{\mu}\partial^{\mu}-{m}_{B}+g_{\sigma B}\sigma-g_{\omega B}\gamma_{\mu}\omega^{\mu}
\nonumber\\
&&-\frac{1}{2}g_{\rho B}\gamma_{\mu}\tau\cdot\rho^{\mu})\Psi_{B}+\frac{1}{2}\left(\partial_{\mu}\sigma\partial^{\mu}\sigma-m_{\sigma}^{2}\sigma^{2}\right)
\nonumber\\
&&-\frac{1}{4}\omega_{\mu \nu}\omega^{\mu \nu}+\frac{1}{2}m_{\omega}^{2}\omega_{\mu}\omega^{\mu}-\frac{1}{4}\rho_{\mu \nu}\cdot\rho^{\mu \nu}
\nonumber\\
&&+\frac{1}{2}m_{\rho}^{2}\rho_{\mu}\cdot\rho^\mu-\frac{1}{3}g_{2}\sigma^{3}-\frac{1}{4}g_{3}\sigma^{4}
\nonumber\\
&&+\sum_{\lambda=e,\mu}\overline{\Psi}_{\lambda}\left(i\gamma_{\mu}\partial^{\mu}
-m_{\lambda}\right)\Psi_{\lambda}
\nonumber\\
&&+\mathcal{L}^{YY}
.\
\end{eqnarray}
The last term representing the contribution of the mesons $\sigma^{*}$ and $\phi$ reads
\begin{eqnarray}
\mathcal{L}^{YY}&=&\sum_{B}g_{\sigma^{*} B}\overline{\Psi}_{B}\Psi_{B}\sigma^{*}-\sum_{B}g_{\phi B}\overline{\Psi}_{B}\gamma_{\mu}\Psi_{B}\phi^{\mu}
\nonumber\\
&&+\frac{1}{2}\left(\partial_{\mu}\sigma^{*}\partial^{\mu}\sigma^{*}-m_{\sigma^{*}}\sigma^{*2}\right)-\frac{1}{4}S_{\mu \nu}S^{\mu \nu}
+\frac{1}{2}m_{\phi}^{2}\phi_{\mu}\phi^{\mu}
.\
\end{eqnarray}
Here, $S_{\mu \nu}=\partial_{\mu}\phi_{\nu}-\partial_{\nu}\phi_{\mu}$. Then the RMF approximation is used~\cite{Glendenning97}.

The condition of $\beta$ equilibrium in neutron star matter demands the chemical equilibrium
\begin{eqnarray}
\mu_{i}=b_{i}\mu_{n}-q_{i}\mu_{e},
\end{eqnarray}
where $b_{i}$ is the baryon number of a species $i$.

In this work, we choose the nucleon coupling constant GL85 set~\cite{Glendenning85}:
the saturation density $\rho_{0}$=0.145 fm$^{-3}$, binding energy B/A=15.95 MeV, a compression modulus $K=285$ MeV, charge symmetry coefficient $a_{sym}$=36.8 MeV and the effective mass $m^{*}/m$=0.77.

For the hyperon coupling constant, we define the ratios: $x_{\sigma h}=\frac{g_{\sigma h}}{g_{\sigma n}}$, $x_{\omega h}=\frac{g_{\omega h}}{g_{\omega n}}$ and $x_{\rho h}=\frac{g_{\rho h}}{g_{\rho n}}$. Here, $h$ denotes hyperons $\Lambda, \Sigma$ and $\Xi$ and $n$ nucleons.

We choose $x_{\rho \Lambda}=0, x_{\rho \Sigma}=2, x_{\rho \Xi}=1$ by quark constituent SU(6) symmetry~\cite{Dover84,Schaffner96}.
The ratio of hyperon coupling constant to nucleon coupling constant is determined in the range of $\sim$ 1/3 to 1~\cite{Glendenning91}.
So we choose $x_{\sigma \Lambda}$=0.4, 0.5, 0.6, 0.7, 0.8, 0.9,1.0. For each $x_{\sigma \Lambda}$, the $x_{\omega \Lambda}$ should fit to the hyperon well depth~\cite{Glendenning97}

\begin{eqnarray}
U_{h}^{(N)}=m_{n}\left(\frac{m_{n}^{*}}{m_{n}}-1\right)x_{\sigma h}+\left(\frac{g_{\omega n}}{m_{\omega}}\right)^{2}\rho_{0}x_{\omega h}
.\
\end{eqnarray}
Similarly, we also choose $x_{\sigma \Sigma}, x_{\sigma \Xi}$=0.4, 0.5, 0.6, 0.7, 0.8, 0.9,1.0 and for each $x_{\sigma \Sigma}, x_{\sigma \Xi}$ we choose $x_{\omega \Sigma}, x_{\omega \Xi}$ to fit to the hyperon well depth.

The experiments show $U_{\Lambda}^{(N)}=-30$ MeV~\cite{Batty97},
$ U_{\Sigma}^{(N)}=10\sim40$ MeV~\cite{Kohno06,Harada05,Harada06,Friedman07} and $U_{\Xi}^{(N)}=-28$ MeV~\cite{Schaffner-Bielich00}. Then we can choose $U_{\Lambda}^{(N)}=-30$ MeV, $ U_{\Sigma}^{(N)}$=+40 MeV and $U_{\Xi}^{(N)}=-28$ MeV.

The parameters that fit to the experimental data of the hyperon well depth are listed in Table~\ref{tab2}.

\begin{table}[!htbp]
\centering
\caption{The hyperon coupling constants fitting to the experimental data of the well depth, which are $U_{\Lambda}^{N}=-30$ MeV, $U_{\Sigma}^{N}=+40$ MeV and $U_{\Xi}^{N}=-28$ MeV, respectively. For the positive $ U_{\Sigma}^{(N)}$ will restrict the production of the hyperon $\Sigma$ we can only choose $x_{\sigma \Sigma}=0.4$ and $x_{\omega \Sigma}$=0.825 while $x_{\sigma \Sigma}=0.5$ and $x_{\omega \Sigma}$=0.9660 can be deleted.}
\label{tab2}
\begin{tabular}[t]{lllllllll}
\hline\noalign{\smallskip}
$x_{\sigma \Lambda}$ &$x_{\omega \Lambda}$&$U_{\Lambda}^{(N)}$&$x_{\sigma \Sigma}$  &$x_{\omega \Sigma}$ &$U_{\Sigma}^{(N)}$&$x_{\sigma \Xi}$     &$x_{\omega \Xi}$    &$U_{\Xi}^{(N)}$\\
\hline
0.4               &0.3679              &-30.0300    &0.4               &0.8250              &40.0005 &0.4               &0.3811              &  -28.0041\\
0.5               &0.5090              &-30.0100    &\underline{0.5}   &\underline{0.9660}  &40.0044&0.5               &0.5221              &  -28.0002  \\
0.6               &0.6500              &-30.0032    &&& &0.6               &0.6630              & -28.0116\\
0.7               &0.7909              &-30.0146    &&& &0.7               &0.8040              & -28.0076\\
0.8               &0.9319              &-30.0106    &&& &0.8               &0.9450              &-28.0037\\
\noalign{\smallskip}\hline
\end{tabular}
\vspace*{0.7cm}  
\end{table}

As the positive $ U_{\Sigma}^{(N)}$ will restrict the production of the hyperon $\Sigma$ we can only choose $x_{\sigma \Sigma}=0.4$ and $x_{\omega \Sigma}$=0.825 while $x_{\sigma \Sigma}=0.5$ and $x_{\omega \Sigma}$=0.9660 can be deleted.

From the parameters in Table~\ref{tab2} we can make up 25 sets of suitable parameters (named as NO.01, NO.02, ..., NO.25, respectively), for which set we calculate the mass of the neutron star by the Oppenheimer-Volkoff (O-V) equation~\cite{Glendenning97}

\begin{eqnarray}
\frac{\mathrm dp}{\mathrm dr}&=&-\frac{\left(p+\varepsilon\right)\left(M+4\pi r^{3}p\right)}{r \left(r-2M \right)}
,\\\
M&=&4\pi\int_{0}^{r}\varepsilon r^{2}\mathrm dr
.\
\end{eqnarray}

\begin{figure}[!htp]
\begin{center}
\includegraphics[width=3.5in]{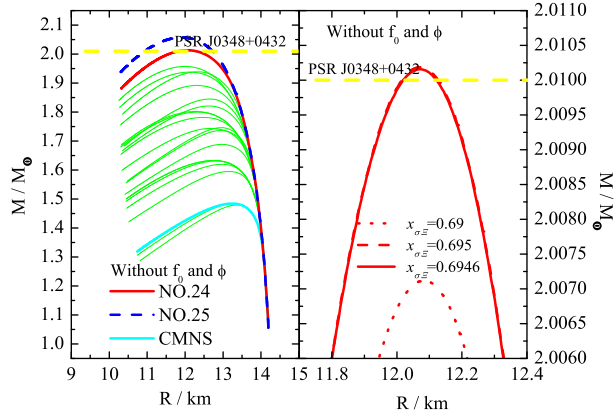}
\caption{The mass as a function of the radius of the neutron star. Here, the mesons $\sigma^{*}$ and $\phi$ are not considered.}
\label{fig1}
\end{center}
\end{figure}

Figure~\ref{fig1} shows the neutron star mass as a function of the radius without considering the contribution of the mesons $\sigma^{*}$ and $\phi$. We see only the parameter NO.24 ( $x_{\sigma \Lambda}=0.8, x_{\omega \Lambda}$=0.9319; $x_{\sigma \Sigma}=0.4, x_{\omega \Sigma}=0.825$; $x_{\sigma \Xi}=0.7, x_{\omega \Xi}=0.804$. The maximum mass calculated is M$_{max}$=2.0132 M$_{\odot}$ ) and NO.25 ( $x_{\sigma \Lambda}=0.8, x_{\omega \Lambda}$=0.9319; $x_{\sigma \Sigma}=0.4, x_{\omega \Sigma}=0.825$; $x_{\sigma \Xi}=0.8, x_{\omega \Xi}=0.945$. The maximum mass calculated is M$_{max}$=2.0572 M$_{\odot}$ ) can give the mass greater than that of the massive neutron star PSR J0348+0432 ( 2.01 M$_{\odot}$ ). Since the mass corresponding to NO.24 is closer to the mass 2.01 M$_{\odot}$, we have fine tuning parameters from $x_{\sigma \Xi}$=0.7 to 0.69, 0.695, 0.6946, respectively. And the $x_{\omega \Xi}$ obtained by fitting to the hyperon well depth. Thus, we get one set of parameters ( $x_{\sigma \Lambda}=0.8, x_{\omega \Lambda}$=0.9319; $x_{\sigma \Sigma}=0.4, x_{\omega \Sigma}=0.825$; $x_{\sigma \Xi}=0.6946, x_{\omega \Xi}=0.7964$ ) corresponding to the mass of the massive neutron star PSR J0348+0432 ( 2.01 M$_{\odot}$ ). In the next step, we use this set of parameters to study whether the the massive neutron star PSR J0348+0432 can change into a hyperon star.

\section{The massive neutron star PSR J0348+0432 can be a hyperon star without considering the mesons $\sigma^{*}$ and $\phi$}

The hyperon star transition density $\rho_{t}$ of a neutron star is defined as the baryon number density at which the hyperons produce.

\begin{figure}[!htp]
\begin{center}
\includegraphics[width=3in]{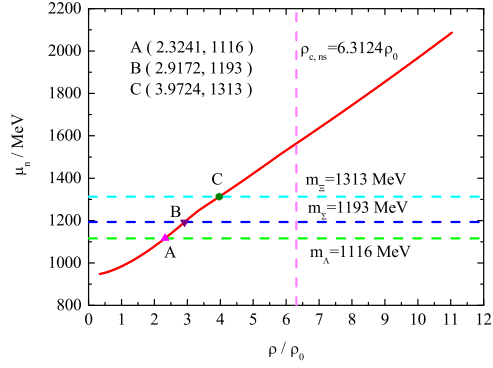}
\caption{The chemical potentials of the nucleons as a function of the baryon number density. The mesons $\sigma^{*}$ and $\phi$ are not considered.}
\label{fig2}
\end{center}
\end{figure}

The chemical potentials of the nucleons as a function of the baryon number density is given in Fig.~\ref{fig2}. The baryon number density is in units of the density of ordinary nuclear matter $\rho_{0}$=0.145 fm$^{-3}$.

We see the chemical potentials of the nucleons increases with the increase of the baryon number density. As the chemical potentials of the nucleons exceed the mass of the $\Lambda$, i.e. 1116 MeV, the $\Lambda$s produce. Afterwards, as the nucleons's chemical potentials exceed the mass of the $\Xi$, i.e. 1313 MeV, the $\Xi$s produce. For the positive $ U_{\Sigma}^{(N)}$ restricts the production of the hyperon $\Sigma$, even the nucleons's chemical potentials exceed the mass of the $\Sigma$, i.e. 1193 MeV, the $\Sigma$ would not appear. We also can see that the central baryon number density is $\rho_{c}$=6.3142 $\rho_{0}$ and within the neutron star the hyperons $\Lambda$ and $\Xi$ all will produce.

So, within the neutron star PSR J0348+0432, with the increase of the baryon number density more and more nucleons will decay to hyperons $\Lambda$ and $\Xi$. At the center of the neutron star, the relative number density of
\setcounter{page}{176}
the hyperons will arrive at its maximum value and that of the neutrons will arrive at its minimum value.

\begin{figure}[!htp]
\begin{center}
\includegraphics[width=3in]{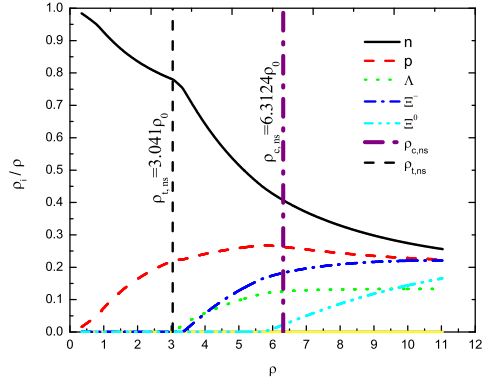}
\caption{The relative number density as a function of the baryon number density. Here, the maseons $\sigma^{*}$ and $\phi$ are not considered.}
\label{fig3}
\end{center}
\end{figure}

Figure~\ref{fig3} displays the relative particle number density of n, p, $\Lambda$, $\Xi^{-}$ and $\Xi^{0}$ as a function of the baryon number density. We see that the firstly generated hyperon is $\Lambda$, which will produce as $\rho\geq$3.041$\rho_{0}$ (at which the relative number density of $\Lambda$ is $\rho_{\Lambda}/\rho$=0.001358\%). So the massive neutron star PSR J0348+0432 can be a hyperon star in this case. The hyperon star transition density of the neutron star is $\rho_{t,ns}$=3.041$\rho_{0}$ as the mesons $\sigma^{*}$ and $\phi$ not being considered.

\section{The massive neutron star PSR J0348+0432 also can be a hyperon star with the mesons $\sigma^{*}$ and $\phi$ being considered}

\begin{figure}[!htp]
\begin{center}
\includegraphics[width=3in]{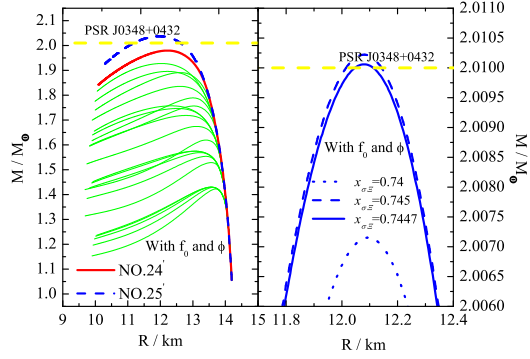}
\caption{The mass as a function of the radius of the neutron star. In this case, the mesons $\sigma^{*}$ and $\phi$ are considered.}
\label{fig4}
\end{center}
\end{figure}

Adopting the similar steps as above, we can obtain a model of the massive neutron star PSR J0348+0432 as the mesons $\sigma^{*}$ and $\phi$ being considered: $x_{\sigma \Lambda}=0.8, x_{\omega \Lambda}$=0.9319; $x_{\sigma \Sigma}=0.4, x_{\omega \Sigma}=0.825$; $x_{\sigma \Xi}=0.7447, x_{\omega \Xi}=0.8671$. This case can be seen in Fig.~\ref{fig4}.

\begin{figure}[!htp]
\begin{center}
\includegraphics[width=3in]{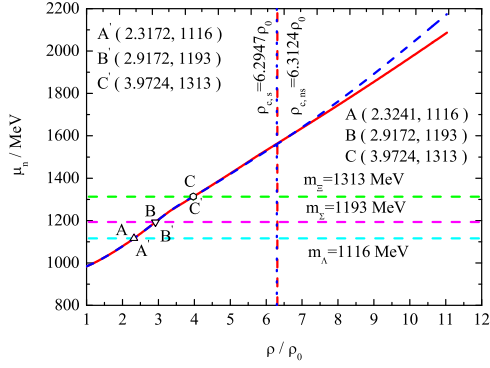}
\caption{The chemical potentials of the nucleons as a function of the baryon number density. The blue curve include the mesons $\sigma^{*}$ and $\phi$ and the red curve do not consider them.}
\label{fig5}
\end{center}
\end{figure}

Figure~\ref{fig5} give the chemical potentials of the nucleons as a function of the baryon number density. The blue curve include the mesons $\sigma^{*}$ and $\phi$ and the red curve do not consider them. We see the chemical potentials of the nucleons without  considering the mesons $\sigma^{*}$ and $\phi$ is almost the same as that with considering them since the baryon number density is less than the central one {\color{blue}$\rho<\rho_{c,ns}\approx\rho_{c,s}$}. So, the hyperons $\Lambda$s and $\Xi$s would also produce with the increase of the baryon number density. Based on the same reason above, the hyperon $\Sigma$ also can not appear.

\begin{figure}[!htp]
\begin{center}
\includegraphics[width=3in]{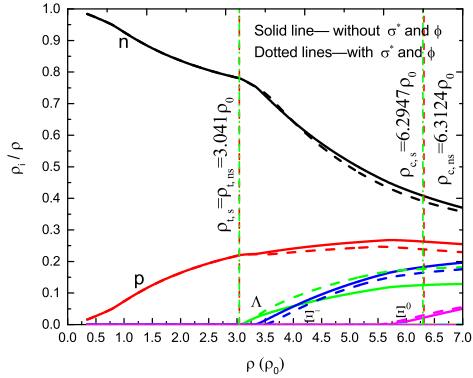}
\caption{The relative number density as a function of the baryon number density. Here, the blue curve include the mesons $\sigma^{*}$ and $\phi$ while the red curve do not include them.}
\label{fig6}
\end{center}
\end{figure}

The relative particle number density as a function of the baryon number density with and without considering the mesons $\sigma^{*}$ and $\phi$ are shown in Fig.~\ref{fig6}. We see, considering the contribution of the mesons $\sigma^{*}$ and $\phi$, the hyperon $\Lambda$ produces as $\rho$=3.041 $\rho_{0}$, at which the relative number density of $\Lambda$ is $\rho_{\Lambda}/\rho$=0.001374\%. That is to say, considering the mesons $\sigma^{*}$ and $\phi$ the massive neutron star PSR J0348+0432 also can be a hyperon star. The hyperon star transition density of the neutron star also is $\rho_{t,s}$=3.041$\rho_{0}$ as the mesons $\sigma^{*}$ and $\phi$ being considered.

\section{The canonical mass neutron star also can be a hyperon star as the mesons $\sigma^{*}$ and $\phi$ not being considered}
\begin{figure}[!htp]
\begin{center}
\includegraphics[width=3in]{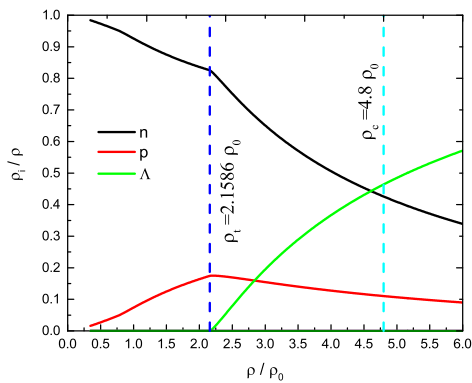}
\caption{The hyperon star transition density of the canonical mass neutron star with the mesons $\sigma^{*}$ and $\phi$ not being considered.}
\label{fig7}
\end{center}
\end{figure}

In our calculations, without considering the mesons $\sigma^{*}$ and $\phi$ the parameters $x_{\sigma \Lambda}=0.4, x_{\omega \Lambda}$=0.3679; $x_{\sigma \Sigma}=0.4, x_{\omega \Sigma}=0.825$; $x_{\sigma \Xi}=0.8, x_{\omega \Xi}=0.9450$ give a model of a canonical mass neutron star ( denoted as CMNS, which mass obtained by us is $M$=1.4843 M$_{\odot}$. See Fig.~\ref{fig1} ). Figure~\ref{fig7} shows the hyperon star transition density of the CMNS with the mesons $\sigma^{*}$ and $\phi$ not being considered. We see the central baryon density of the CMNS is $\rho_{c}$=4.8 $\rho_{0}$. As the baryon number density is at $\rho$=2.1586$\rho_{0}$, the $\Lambda$s begin to produce (though its relative number density is very small: only $\rho_{\Lambda}/\rho$=0.00793\%). This means the CMNS can be a hyperon star and the hyperon star transition density of the CMNS is $\rho_{t}$=2.1586$\rho_{0}$.

\section{Summary}
In this paper, whether the massive neutron star PSR J0348+0432 can be a hyperon star is studied in the framework of the RMF theory by choosing the suitable hyperon coupling constants. It is found that the hyperons ($\Lambda$s) will first produce at the baryon number density $\rho$=3.041 $\rho_{0} $ as the mesons $\sigma^{*}$ and $\phi$ not being considered. So the massive neutron star PSR J0348+0432 can be a hyperon star. As the mesons $\sigma^{*}$ and $\phi$ being considered, the first kind of hyperons ($\Lambda$s) also can produce at the baryon number density $\rho$=3.041 $\rho_{0} $. Therefore, as the mesons $\sigma^{*}$ and $\phi$ are considered the massive neutron star PSR J0348+0432 also can {\color{blue}be} a hyperon star. For the two cases, the hyperon star transition density of the neutron star are the same. For the CMNS, as the baryon number density is $\rho$=2.1586 $\rho_{0}$, the first kind of hyperons ($\Lambda$s) produce as the mesons $\sigma^{*}$ and $\phi$ not being considered. Then the hyperon star transition density of the CMNS is $\rho_{t}$=2.1548 $\rho_{0}$, which is smaller than that of the massive neutron star PSR J0348+0432.

In order to reconcile the existence of massive neutron stars as the massive neutron star PSR J0348+0432 with the softening of hyperon equations of state due to the increasing number of degrees of freedom, we choose the larger coupling constants $x_{\omega h}$, which can provide larger repulsion force. About this case, a lot of works have been done on it over the last years. Our results confirms the conclusions obtained by Bednarek et al~\cite{Bednarek12}, Kolomeitsev et al~\cite{Kolomeitsev05}, Dexheimer et al~\cite{Dexheimer08} and Weissenborn et al~\cite{Weissenbornprc12}.

\section*{Acknowledgements}

We are thankful to anonymous referee for valuable suggestions and to Prof Shan-Gui Zhou for fruitful discussions during my visit to the Institute of Theoretical Physics, Chinese Academy of Sciences. This work was supported by the Natural Science Foundation of China ( Grant No. 11447003 ) and the Scientific Research Foundation of the Higher Education Institutions of Anhui Province, China ( Grant No. KJ2014A182 ).


\end{document}